\documentclass{iau} 
\usepackage{graphicx}

\def\msun{\hbox{M$_\odot$}}

\def\mstar{\hbox{M$_\star$}}
\def\t4{\hbox{t$_{\rm 4}$}}

\def\relation{\hbox{SFR vs. M$_{V}^{\rm brigh}$}}

\def\cm3{\hbox{cm$^{-3}$}}

\def\ga{\hbox{$\Gamma$}}

\newcommand{\araa}{ARA\&A}
\newcommand{\apj}{ApJ}
\newcommand{\aj}{AJ}
\newcommand{\mnras}{MNRAS}
\newcommand{\aap}{A\&A}

\newcommand{\nat}{Nature}

\newcommand{\apjs}{ApJS}

\title[The imprints of the galactic environment on YSC populations] 
{The Imprints Of Galactic Environment \\ On Cluster Formation and Evolution}

\author[A. Adamo]   
{Angela Adamo$^1$\\
}

\affiliation{$^1$ Department of Astronomy, Oskar Klein Centre, Stockholm University, AlbaNova University Centre, SE-106 91 Stockholm, Sweden \\ email: {\tt adamo@astro.su.se} \\[\affilskip]}

\pubyear{2015}
\volume{316}  
\jname{Formation, evolution, and survival of massive star clusters}
\editors{C. Charbonnel \& A. Nota, eds.}
\begin{document}

\maketitle

\begin{abstract}
Young star clusters (YSCs) appear to be a ubiquitous product of star formation in local galaxies, thus, they can be used to study the star formation process at work in their host galaxies. Moreover, YSCs are intrinsically brighter that single stars, potentially becoming the most important tracers of the recent star formation history in galaxies in the local Universe. In local galaxies, we also witness the presence of a large population of evolved star clusters, commonly called globular clusters (GCs). GCs peak formation history is very close to the redshift ($z\sim$2) when the cosmic star formation history reached the maximum. Therefore, GCs are usually associated to extreme star formation episodes in high-redshift galaxies. It is yet not clear whether YSCs and GCs share a similar formation process (same physics under different interstellar medium conditions) and evolution process, and whether the former can be used as progenitor analogs of the latter. In this invited contribution, I review general properties of YSC populations in local galaxies. I will summarise some of the current open questions in the field, with particular emphasis to whether or not galactic environments, where YSCs form, leave imprints on the nested populations. The importance of this rapidly developing field can be crucial in understanding GC formation and possibly the galactic environment condition where this ancient population formed. 
\end{abstract}

\firstsection 
\section{Introduction}

YSCs are usually identified as gravitationally bound stellar structures, with radii between 0.5 up to several parsecs and masses between $10^3$ and $10^7$ \msun (\cite[Portegies Zwart et al. 2010]{2010ARA&A..48..431P}).
Thanks to the HST spatial resolution and multiband coverage the YSC field has been continuously advancing. 

YSCs are easily observed in local galaxies which are able to sustain star formation. This is not always true at the lowest star-forming regimes. For example, in dwarf galaxies, star and cluster formation appears dominated by stochastic processes (\cite[Cook et al. 2012]{2012ApJ...751..100C}). The range of star formation rate (SFR) and star formation rate density ($\Sigma_{SFR}$) of dwarf galaxies hosting one or few YSCs or none overlaps significantly (e.g.  \cite[Adamo \& Bastian 2015]{aa..nb..2015Spr}). On the other hand, some local dwarf galaxies are able to form very massive YSCs, in some cases even more massive than usually observed in spiral galaxies like the Milky Way (e.g. \cite[Billett et al. 2002]{2002AJ....123.1454B}). This way of forming massive YSCs has been associated to the possibility that dwarf galaxies at high $z$ are the places where low metallicity GCs have formed (\cite[Elmegreen et al. 2012]{2012ApJ...757....9E}). However, whether cluster formation in dwarf galaxies is dominated by stochastic events or properties of the galactic environment is still an open question.

YSC formation in spiral galaxies is a continuos process, which reflects the fact that star formation has proceeded with a constant pace for at least several hundreds of Myr. This equilibrium is the result of a regulation process between gas feeding the disk system, the dynamics, and the feedback produced by star formation. Typically spiral arms are the preferential places where star formation takes place in disk galaxies. Similarly to "traffic jams", i.e. regions where rotation speeds slows down, giant molecular clouds (GMCs) entering the spiral arm region can finally grow in mass. Once GMCs move out of the "jam" they will fragment and collapse forming stars and star clusters (e.g. \cite[Elmegreen 2011, Dobbs 2014]{2011EAS....51...19E, 2014IAUS..298..221D}). However, if spiral arms are visually the most dominating features in disk galaxies, dynamically, the presence of bars can also impact star formation often leading to the formation of star-forming circumnuclear rings. In galaxies like the Milky Way, the end of the bar and the central circumnuclear ring are preferential places where massive stars and YSCs form (e.g. \cite[Davies et al. 2012, Stolte 2013]{2012MNRAS.419.1860D, 2013msao.confE..30S}). It is also important to keep in mind that the presence of a large amount of molecular gas is not always a guarantee of constantly efficient star and cluster formation. For example, the central molecular zone of our Milky Way is likely to alternate episodic starburst events to phases of inefficient star formation (e.g. \cite[Krumholz \& Kruijssen 2015]{2015MNRAS.453..739K}).

Finally, a preferential channel to form numerous and massive YSC populations is during starburst episodes. During a starburst the SFR in the galaxy increases to levels which are impossible to sustain for a Hubble time. In the local Universe, starbursts are usually associated to interaction and merging processes between galaxies, with at least one of the systems involved being gas rich. This is not necessarily true in dwarf galaxies, where star formation appear to be dominated by episodic and, sometimes, extremely efficient star formation processes. This is the case of nearby dwarf starburst galaxies like NGC1705 (\cite[Annibali et al. 2009]{2009AJ....138..169A}); NGC 4449 (\cite[Annibali et al. 2011]{2011AJ....142..129A}), NGC5253 (\cite[Calzetti et al. 2015b]{2015ApJ...811...75C}), among many others. Numerous YSC populations, with cluster masses reaching up to a few time $10^7$ \msun\, have been easily found in merging systems like NGC7252 and NGC1316 (\cite[Bastian et al. 2006]{2006A&A...448..881B}), and ongoing mergers like the Antennae (\cite[Whitmore et al. 2010]{2010AJ....140...75W}). 

At this point, the reader may ask the question whether cluster populations forming in different galaxies share any similarity at all. Another valid question could be related to whether, given a YSC population with certain properties, it is possible to reconstruct the galactic environment where these clusters are forming. As I will summarise in this review, YSC populations forming in the nearby Universe share a large spectrum of statistical properties. There is an underlying universality in the properties that can describe YSC populations in local galaxies. However, several studies, in the last decay, also show that the cluster population of each galaxy carry some imprints that are tightly correlated to the physical conditions of the gas in their host galaxies. Potentially, YSC properties become a key to understand both how star formation proceeds and the undergone formation history of the host galaxy. 

\section{YSC mass and luminosity function}
After almost 3 decay of HST based studies of YSC populations in local galaxies it is fairly accepted that the YSC luminosity function can be described by a power law function, $dN\propto L^{\alpha}$, with $\alpha \sim-2$. This distribution is the result of an underlying power-law distribution of the YSC masses with a similar index. Similar slopes have been observed for the luminosity distribution of increasing stellar aggregate boundaries (\cite[Elmegreen et al.2006]{2006ApJ...644..879E}), confirming that cluster formation is well nested inside the hierarchical star formation process driven by turbulence. 

However, this picture is not complete. In the literature, a quick survey of the recovered power-law indexes as a function of the luminosity range above which clusters luminosity are distributed can easily show that the observed $\alpha$ values variate from $-1.6$ to $-2.8$ (see Figure 12 by \cite[Annibali et al. 2011]{2011AJ....142..129A}). In particular the steepening of the slopes seams to increase at the brightest luminosity bins (\cite[Gieles 2010]{2010ASPC..423..123G}), suggesting that we observe less luminous clusters than expected from a power-law function of index $-2$. 

Another way to address this problem is to look at the underlying YSC mass functions. Does the observed YSC mass function shows a dearth of massive star clusters? Or in other words, is the YSC mass function a power-law with no upper-mass limits or does it have some smooth truncation? The observed steepening of the YSC luminosity function is easily reproducible if the underlying YSC mass function has the shape of a Schechter function, i.e. a power-law function with index $-2$ and an exponential decline above a certain truncation mass, \mstar\, (\cite[Gieles et al. 2006]{2006A&A...446L...9G}). Using statistical arguments, \cite[Larsen(2009)]{2009A&A...494..539L} noticed that if the YSC mass function of the Milky Way (as prototype of local spiral galaxies) would be a pure power-law function with no upper limits, then our own Galaxy should have been able to form in the last few Gyr clusters with masses $\sim 10^6$ \msun. A Schechter mass function with \mstar$\sim 2\times10^5$ \msun\, can easily explain the dearth of such massive YSCs in local spirals. The presence of a truncation is not in contrast with the very massive YSCs found in starburst galaxies. A fit to the YSC mass function of the Antennae shows that a Schechter function with \mstar$\sim 10^6$ \msun\, can reproduce the observed distribution of cluster masses. Therefore, the truncation mass doesn't appear to be a universal value but it is more likely linked to the galactic environment where the cluster population is forming (\cite[Larsen 2009]{2009A&A...494..539L}).

\begin{figure}[b]
\begin{center}
 \includegraphics[width=2.9in]{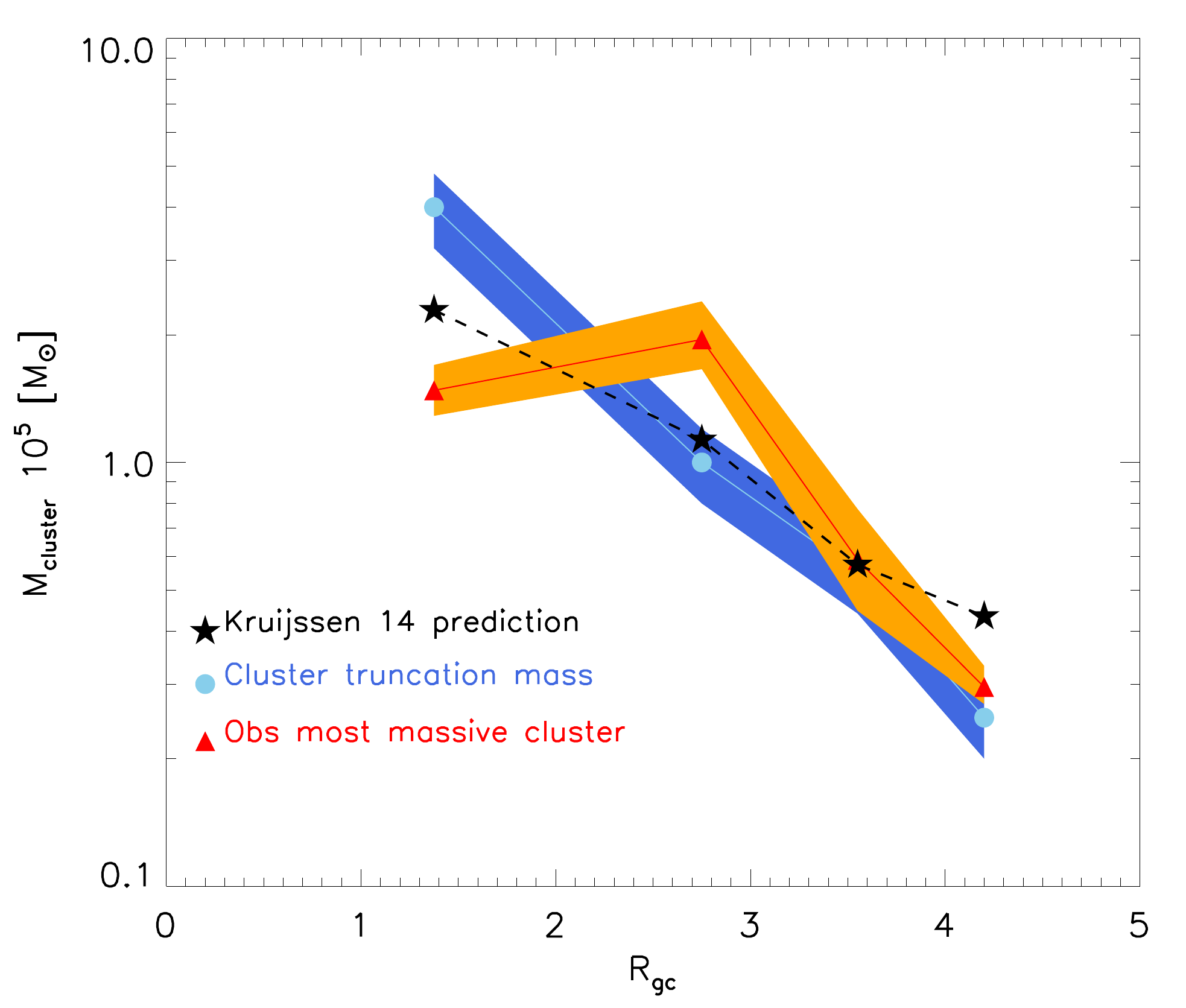} 
 \includegraphics[width=2.3in]{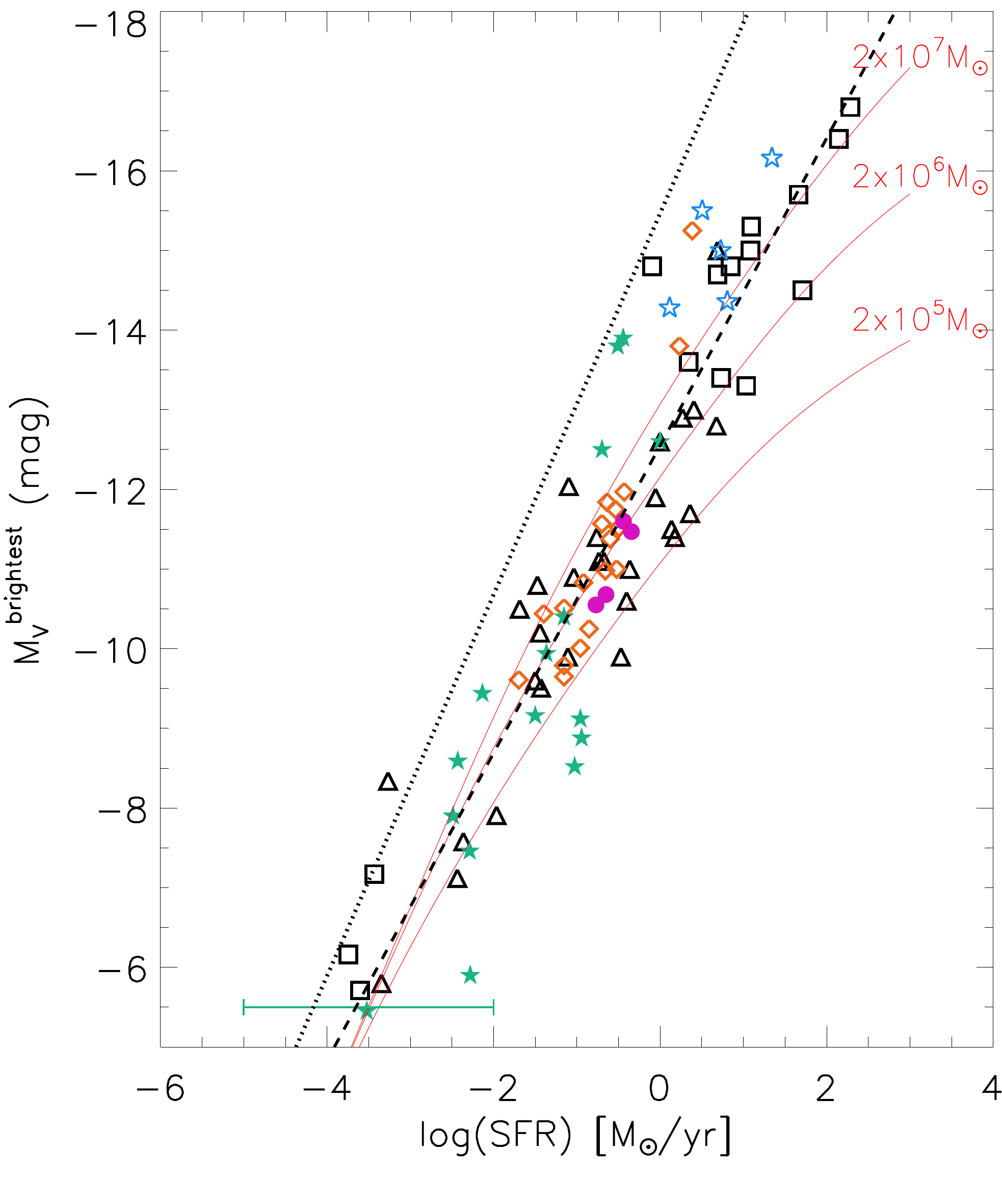} 
 \caption{{\bf Left}: The observed mass of the most massive cluster (red tringles), the recovered \mstar (cyan dots), and the estimated maximum cluster mass (black stars) using models by \cite[Kruijssen (2014)]{2014CQGra..31x4006K} found in annuli of increasing galactocentric distances (kpc) in M83. The shadowed regions show the uncertainties. {\bf Right}: The \relation\, relation. A full description of the compiled data is accessible in \cite[Adamo et al. (2015)]{2015MNRAS.452..246A}. The dashed line is the fit produced by \cite[Larsen(2002)]{2002AJ....124.1393L} to the spiral sample showed with triangle symbols. The green horizontal bar is the range covered in SFR by dwarf galaxies which do not host any bound YSC formed in the last 100 Myr \cite[Cook et al. 2012]{2012ApJ...751..100C}. The dotted line shows the region where galaxies would have fallen  if 100\% of their star formation is happening in YSCs. The observed trend is explained if about 8\% of the SFR in galaxy is happening in bound YSCs \cite[Bastian 2008]{2008MNRAS.390..759B}. Red solid lines are the regions where galaxies should fall if their mass functions are Schechter functions with increasing \mstar\, (see values in the insets of the plot). Figures from \cite[Adamo et al. (2015)]{2015MNRAS.452..246A}.}
   \label{fig1}
\end{center}
\end{figure}

A similar conclusion has been reached studying the YSC population in the disk of the nearby spiral galaxy M83 (\cite[Adamo et al. 2015]{2015MNRAS.452..246A}). The YSC population of M83 is better reproduced by a Schechter mass function with \mstar$\sim 2\times 10^5$ \msun. A Schechter function is also favoured when analysing the mass function of the M83 YSC population as a function of galactocentric distances. The mass of the most massive cluster found in each bin oscillates around the recovered \mstar\, value, well in agreement with the probabilistic meaning of a truncation mass in the framework of the Schechter function. The change in the \mstar\, is possibly caused by the change in the ISM physical properties (e.g. gas pressure, density) but further investigation is required.

While our understanding of the driving physical parameter(s) that causes \mstar\, to change as a function of galactic environment is not well constrained yet, further evidence coming from studies of the mass functions of globular cluster populations reinforce these findings. \cite[Jord{\'a}n et al. (2007)]{2007ApJS..171..101J} analysed the mass function of the GC populations detected in roughly 90 galaxy members of the Virgo cluster. They report that an evolved Schechter function, parametrised to take into account the mass loss of the population, produces a better fit to the observed mass distributions. In particular, the recovered \mstar\, declines as a function of the total $B$ band luminosity of the galaxy (used as a proxy of the total stellar mass of the galaxy). The decline cannot be explained by the dynamical effects, like friction, and it is probably linked to the properties of the galaxy at the time of the GC formation.

The similarity in the observed properties of YSC and GC population mass functions, although on significantly different mass scales, possibly suggests that the two populations may share the same formation mechanisms under likely different ISM conditions.

\section{YSC scaling relations}

\begin{figure}[b]
\begin{center}
 \includegraphics[width=2.65in]{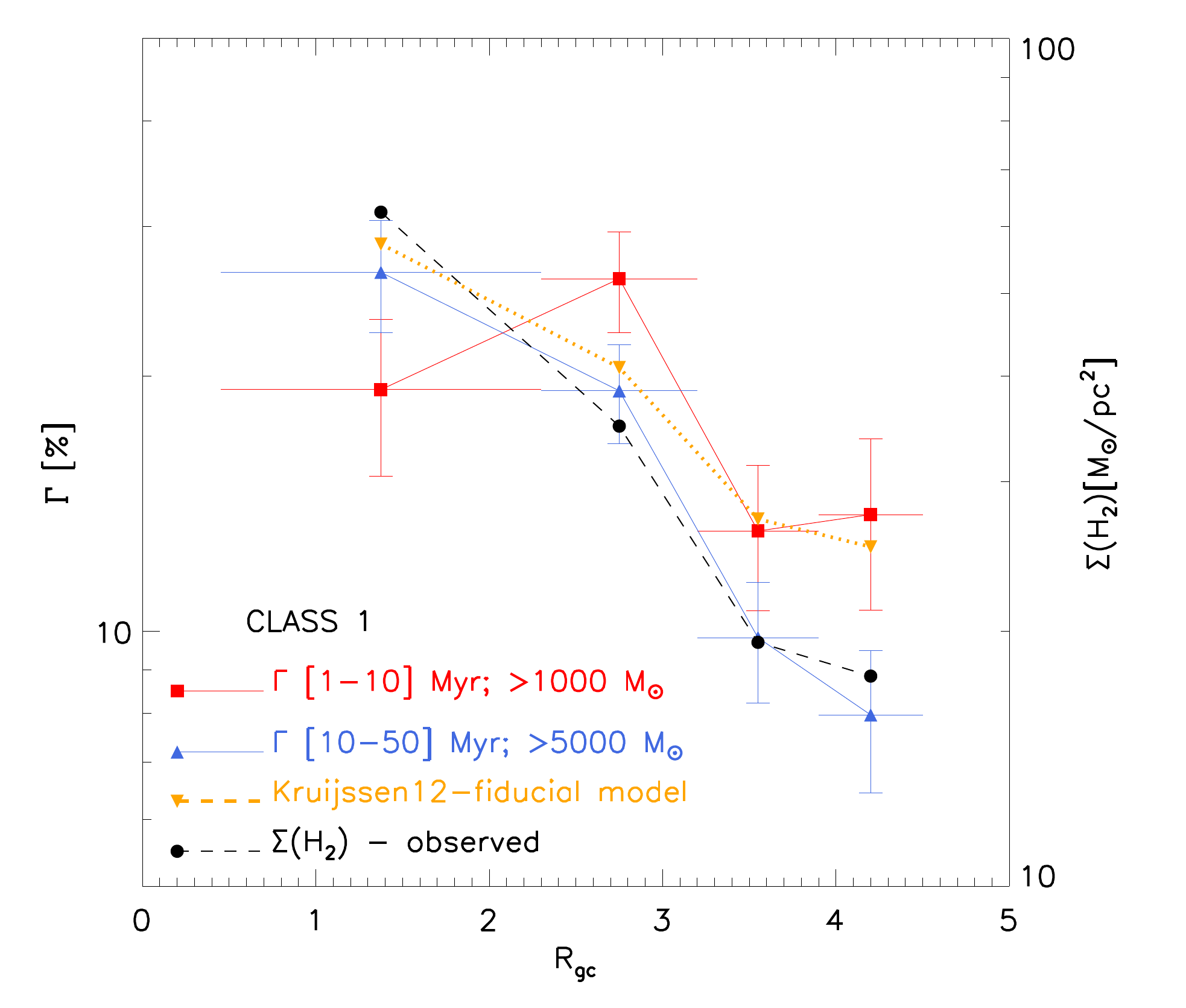} 
 \includegraphics[width=2.6in]{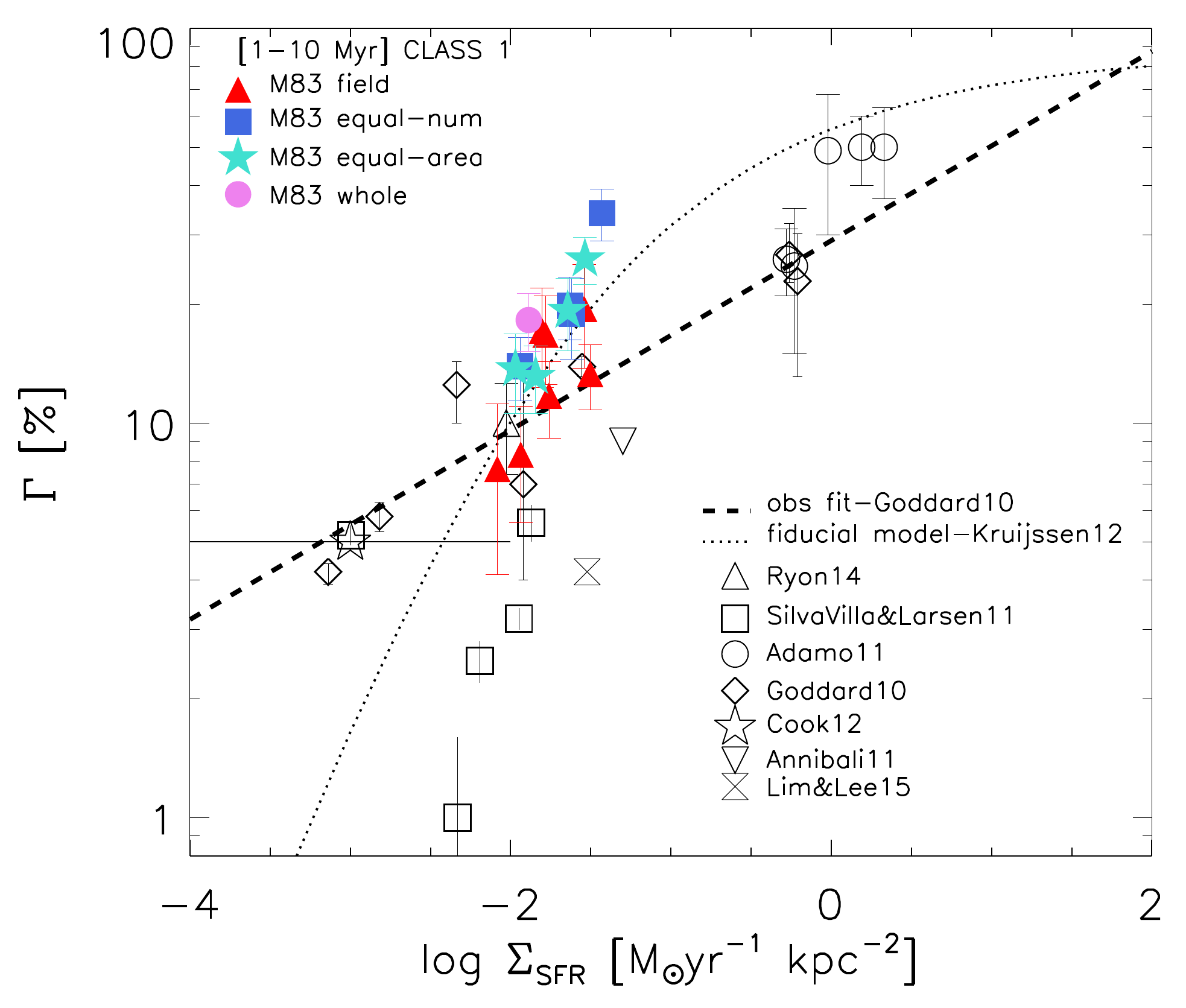} 
 \caption{{\bf Left}: Observed cluster formation efficiency recovered in bins of increasing galactocentric distance. The values have been derived for two different cluster age ranges and SFR indicators (red squares and blue triangles, see inset in the panel). The yellow upside-down triangles show the prediction made with the fiducial model proposed by \cite[Kruijssen(2012)]{2012MNRAS.426.3008K}. The black dots show the observed average gas surface density (unit scales is showed in the right axes) estimated in the same annuli. {\bf Right}: The \ga\, versus $\Sigma_{\rm SFR}$ relation. The plot includes all the data available in the literature to date (see inset). Figures from \cite[Adamo et al. (2015)]{2015MNRAS.452..246A}.}
   \label{fig2}
\end{center}
\end{figure}

The YSC scaling relations are important tools to constrain and probe the cluster formation process and how cluster population properties are linked to their galactic environments. 

The right panel of Fig.\ref{fig1} shows one of the most well-known relation pointed out in YSC studies, the galactic SFR versus brightest cluster in $V$-band (\relation), in one of the most recent and complete compilation (by \cite[Adamo et al. 2015]{2015MNRAS.452..246A}). This relation was originally formulated by \cite[Larsen (2002)]{2002AJ....124.1393L}. A similar scaling relation has been found if the number of YSCs per galaxy is used instead of the SFR (\cite[Whitmore 2000]{2000astro.ph.12546W}). The ($\log N$) or \relation\, relation tells us that cluster formation is primarily a stochastic process, and size-of-sample drives the likelihood of forming very massive clusters. Higher is the SFR in the host galaxy and more numerous clusters are formed, therefore, it is more likely to sample the YSC mass function at the high mass bins and form massive (luminous) clusters. In a recent revisitation of this relation, \cite[Adamo et al. (2015)]{2015MNRAS.452..246A} use monte carlo simulations of YSC populations of increasing \mstar\, to show that the presence of a truncation in the YSC mass function does not violate the observed trend. As SFR increases so it does the \mstar (see red solid lines in the right pane of Fig.\ref{fig1}).

Another important information hidden in the the \relation\, relation is  the position that galaxies occupy in the diagram. \cite[Bastian (2008)]{2008MNRAS.390..759B} shows using monte carlo simulations of YSC populations that if 100\% of star formation happens in bound clusters, that are subsequently destroyed, the observed datapoints should follow the dotted line reproduced in the plot. To explain the current observations, only a smaller fraction of star formation is happening in bound clusters, on average about 8\%.

It is possible to directly measure the cluster formation efficiency (CFE or \ga), or in other words, the amount of star formation happening in bound clusters. \ga\, is the ratio between the total stellar mass found in clusters over a certain age range divided by the total SFR estimated in the same age range. \cite[Goddard et al. (2010)]{2010MNRAS.405..857G} studying the cluster populations of a sample of nearby galaxies first reported that \ga\, appears to scale with $\Sigma_{SFR}$. The \ga\, vs. $\Sigma_{SFR}$ relation has been observed to hold for different type of galaxies (starbursts, \cite[Adamo et al. 2011]{2011MNRAS.417.1904A}; dwarfs, \cite[Cook et al. 2012]{2012ApJ...751..100C}; etc.) and even on local scales (e.g. the M83 study, \cite[Adamo et al. 2015]{2015MNRAS.452..246A}). In the right panel of Fig.\ref{fig2},  the \ga\, vs. $\Sigma_{SFR}$ relation is shown in one of the most updated compilations available in the literature. The observational fit (dashed line) proposed by Goddard et al. (2010) to the initial set is nowadays replaced with the theoretical model (dotted line) proposed by \cite[Kruijssen (2012)]{2012MNRAS.426.3008K}. The scatter around the model is about 3 sigma, but it can easily been accounted for, if one considers that the many tuneable assumptions are made in the model and systematic and random errors in the observational works. The Kruijssen's model predicts that the formation of bound stellar clusters takes place in the highest-density peaks of a hierarchically structured
ISM. At these high densities, the gas goes through a large number of free-fall times on a short time-scale and
therefore reaches higher star formation efficiency. Therefore, YSCs should form more efficiently at high gas pressures (and hence gas surface densities), because these conditions lead to higher density peaks in the ISM and thus favour bound cluster formation. Indeed, the observed \ga\, vs. $\Sigma_{SFR}$ relation indirectly links the CFE to the gas pressure (and surface density) through the observable quantity of the $\Sigma_{SFR}$, a well know proxy for gas surface density (\cite[Kennicutt \& Evans 2012]{2012ARA&A..50..531K}). The left panel of Fig.\ref{fig2} reproduces the decline observed in \ga\, as a function of galactocentric distances in M83. The decline obtained from two different sets of cluster ages and SFR indicators (see inset in the figure) follows quite closely the drop in gas surface density, $\Sigma(H_2)$, reinforcing the idea that the gas pressure (density) plays an important role in the efficiency of the cluster formation process.

In a recent work, \cite[Chandar et al. (2015)]{2015ApJ...810....1C} try to quantify the importance of environmental effects above some "universal" cluster properties as the power-law distribution of the mass function. They derive \ga\, normalising the observed cluster mass function of 6 galaxies by their respective SFR and plot the recovered residuals as a function of SFR. They suggest that the observed trends are within statistical uncertainties and do not reproduced the expectation of the \cite[Kruijssen (2012)]{2012MNRAS.426.3008K} model and, more in general, the observed \ga\, vs. $\Sigma_{SFR}$ relation. However, this reported disagreement is likely not real  (\cite[Kruijssen \& Bastian 2015]{2015arXiv151103286K}). First of all, as discussed in the previous paragraph, there is a well motivated physical link between \ga\, and the $\Sigma_{SFR}$, the latter being an indirect measurement of the gas surface density and thus gas pressure. For this reason, $\Sigma_{SFR}$ and not the absolute SFR should be used when looking at CFE variations. Secondly, it is important to use cluster catalogues which contains as little as possible contaminations from stellar associations. The effects of using SFR instead of $\Sigma_{SFR}$, and catalogues which are contaminated by associations are thoroughly addressed in the recent work by 
 \cite[Kruijssen \& Bastian (2015)]{2015arXiv151103286K}.

\section{Ongoing surveys and goals for the next few years}
The results discussed so far are based on a fairly small sample of galaxies, studied with similar but not homogeneous approaches. To really quantify the role of galactic environment in shaping the YCS formation process we need to access a statistically large sample of galaxies representative of the star-forming system observed in the local Universe. The recently executed  Hubble Space Telescope treasury program LEGUS (Legacy ExtraGalactic Uv Survey, GO 13364, PI Calzetti) has ben designed to fully sampled in SFR, specific SFR, morphology, total stellar mass, typical galaxies located within 12 Mpc.  The data set of each of the 50 galaxies imaged with the HST contains a homogeneous sampling of 5 UV and optical filters. The access to two UV bands on the left side of the Balmer break and 3 bands on the right part is allowing us to derive cluster properties with tighter constraints on their ages and extinctions values. LEGUS is a laboratory for studying both resolved stellar populations and YSCs (\cite[Calzetti et al. 2015a]{2015AJ....149...51C}). The goal of LEGUS is to revolutionise our knowledge of star formation, which is impossible using a single galaxy, but approachable when a fair sampling of typical local galaxies has been selected. As already described in \cite[Calzetti et al. (2015a)]{2015AJ....149...51C}, and soon presented in a forthcoming paper (Adamo et al in prep.), the LEGUS collaboration is constructing high-quality cluster catalogues for most of the targets. 

\begin{figure}[b]
\begin{center}
 \includegraphics[width=5.2in]{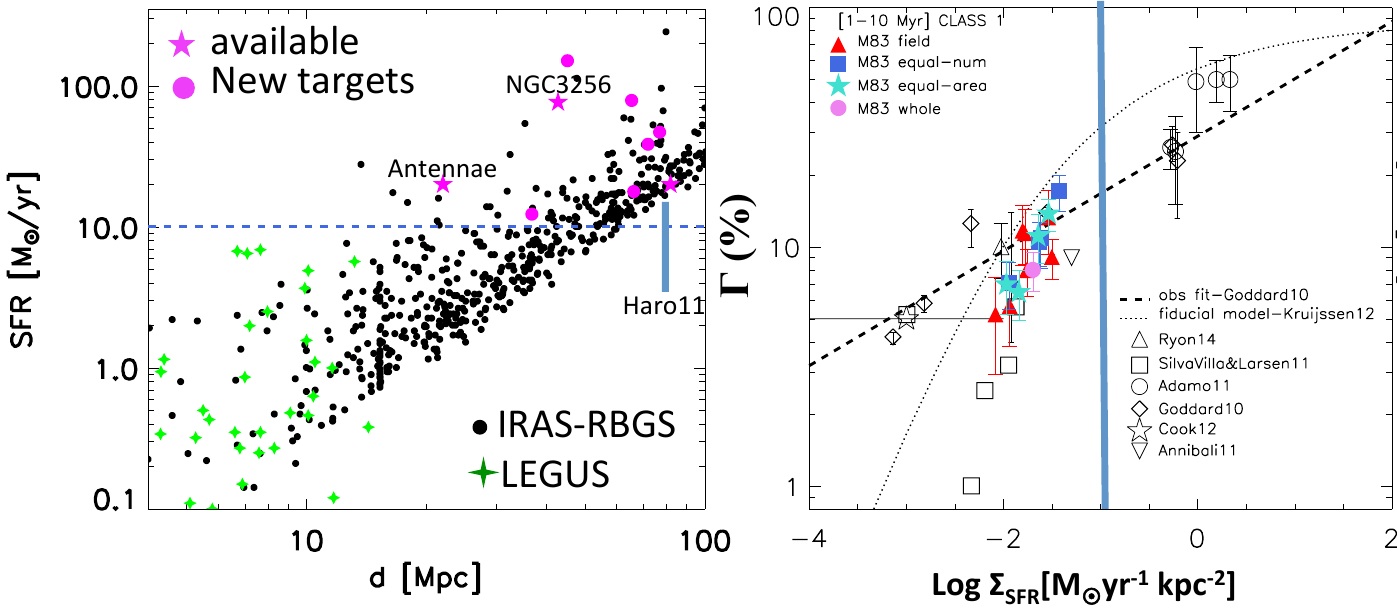} 
 \caption{{\bf Left}: The SFR versus distance space covered by the LEGUS targets (green stars) and Hi-PEEC (magenta stars and dots). The IRAS revised bright galaxy sample is used as a reference (\cite[Sanders et al. 2003]{2003AJ....126.1607S}). {\bf Right}: the \ga\, versus $\Sigma_{SFR}$ diagram. The blue solid line separates the region that will be  probed by LEGUS (left) and by Hi-PEEC (right).}
   \label{fig3}
\end{center}
\end{figure}

Preliminary results produced by the analysis of the cluster population in the inner and outer pointing of the spiral galaxy NGC628 (Adamo et al. in prep.) show that a Schechter function with decreasing \mstar (from the inner to the outer region) is statistically a better representation of the observed YSC mass function in the galaxy. This evidence is also supported by the study of the cluster luminosity function in the two regions. We find that the luminosity function shows a steepening as function of increasing wavelengths and a steepening at the highest luminosity bins in each band. The observed trend is in agreement with the expectations by \cite[Gieles (2010)]{2010ASPC..423..123G}, where the author assumes that the mass function has a truncation and cluster disruption has not yet been significant.

Another upcoming survey, which has been designed to be complementary to the LEGUS one is the Hubble imaging Probe of Extreme Environments and Clusters (Hi-PEEC, PI Adamo). Hi-PEEC (see left panel of Fig.\ref{fig3}) will give us access to star and cluster formation regimes rarely observed in the local Universe but common at high redshift ($z\sim2-3$), when the bulk of the bulk of the GC populations formed. With Hi-PEEC we will be able to probe the properties of the YSC mass function, putting constraints on the environmental changes of \mstar. We will also be able to make a fair comparison with the observed decline in \mstar\, observed in GC populations by \cite[Jord{\'a}n et al. (2007)]{2007ApJS..171..101J}.  Results from LEGUS and Hi-PEEC will be fundamental to yield better constraints on the \ga\, versus $\Sigma_{SFR}$ relation. The right panel of Fig.\ref{fig3} shows which region of the relation the two surveys will be able to probe.

\section{Conclusions}
In this short review, I have mainly summarised recent results about cluster formation. Other key aspects related to cluster evolution such as disruption, size distributions, hierarchical properties of YSCs are not covered here, but recent studies can be found in the literature (e.g., \cite[Adamo \& Bastian 2015, Ryon et al. 2015, Grasha et al. 2015]{aa..nb..2015Spr, 2015MNRAS.452..525R, 2015arXiv151102233G}).

The cluster luminosity and mass functions are powerful tools to understand the role of galactic environment on YSC formations. Recent studies point out that the YCS mass function shows a possible truncation at the high-mass end of the distribution. The truncation seams to change as a function of the galactic environment, but it is not yet understood what galactic physical parameter mainly governs such dependency. Surveys like LEGUS and Hi-PEEC, combined with ancillary data that allow to reconstruct galactic scale properties, like gas surface density, SFR, gas pressure, will produce more stringent constraints on ultimate shape of the YSC mass function.

YSC scaling relations have a key role in linking YSC formation to galaxy evolution, therefore, it is of paramount importance to understand what mechanisms and physical parameters drive these relations.

In a recent numerical work, based on the Illustris cosmological simulations (\cite[Vogelsberger et al. 2014]{2014Natur.509..177V}), \cite[Mistani et al. (2015)]{2015arXiv150900030M} showed that the use of the \ga\, versus $\Sigma_{SFR}$ relation maybe can explain why dwarf ellipticals living in over-dense environments have higher GC specific numbers that dwarf irregular living in the field. Dwarf galaxies that have entered a galaxy cluster have experienced a starburst phase before being stripped of their gas reservoirs. During this phase a larger amount of star clusters have been formed. The analog dwarfs living in the field have never experienced such intense starbursts, therefore, their GC specific number is significantly lower. 

The change in CFE during starburst phases can also have important impacts in driving not only local but also galactic scale feedback. In a recent work by \cite[Bik et al. (2015)]{2015A&A...576L..13B}, the authors find that in the starburst galaxy ESO338, galactic scale ionised cones and outflows can directly be linked to the massive YSC that have recently formed in the galaxy. Since YSCs are favourite places to host massive stars, a CFE of about 50 \%, like in the case of ESO338, will make a more significant impact on the ISM.

A full description of YSC populations in local galaxies requires us to probe both universal properties and galactic environmental dependencies. Recent and upcoming surveys will play a key role in the advancement of this field. The ultimate goal is to build a complete understanding of cluster formation at any redshift and in any galactic environment.

\section*{Acknowledgements}
{\footnotesize I thank the P. L. Lindahls foundation and the Royal Swedesh Academy (KVA) for the provided generous travel fund. N. Bastian and S. Larsen are thanked for their inputs and comments on this draft. I  am very thankful to C. Charbonnel and A. Nota for organising this very interesting symposium.}

\end{document}